\documentclass[12pt,a4paper,english,superscriptaddress,aps,nofootinbib]{revtex4}
\usepackage{amsmath,amssymb,graphicx}
\makeatletter
\usepackage{babel}
\usepackage[active]{srcltx}
\usepackage{graphicx,color}
\usepackage{changebar}
\usepackage{hyperref}
\hypersetup{
	colorlinks=true,
	urlcolor= blue,
	citecolor=blue,
	linkcolor= blue,
	bookmarks=true,
	bookmarksopen=false,}

\usepackage[T1]{fontenc}
\usepackage{ae}
\usepackage[ansinew]{inputenc}
\usepackage{amsmath}
\usepackage{braket}
\usepackage{mathtools}
\usepackage{slashed}
\usepackage{empheq}
\usepackage{tikz}
\usepackage{multirow}
\usetikzlibrary{decorations.pathmorphing}
\usepackage[caption = false]{subfig}

\usepackage{graphicx}
\usepackage{amsfonts}
\usepackage{amsmath}
\newcommand{\sech}{\text{sech}}
\newcommand{\csch}{\text{csch}}

\newcommand{\be}{\begin{equation}}
	\newcommand{\ee}{\end{equation}}
\newcommand{\bea}{\begin{eqnarray}}
	\newcommand{\eea}{\end{eqnarray}}

\begin{document}
	\title{Quantum information entropies for a soliton at hyperbolic well}
	
\author{F. C. E. Lima}
\email[]{E-mail: cleiton.estevao@fisica.ufc.br}
\affiliation{Universidade Federal do Cear\'{a} (UFC), Departamento do F\'{i}sica - Campus do Pici, Fortaleza, CE, C. P. 6030, 60455-760, Brazil.}

\begin{abstract}
In this work, the probability uncertainties related to a stationary quantum system with solitonic mass distribution when subjected to deformable hyperbolic potentials are studied. Shannon's entropy and Fisher's information of a position-dependent mass are calculated. To investigate the concept of Shannon and Fisher entropies of the solitonic mass distribution subject to the hyperbolic potential, it is necessary to obtain the analytic solutions at position and momentum space. For the Hamiltonian operator to be Hermitian, we consider the stationary Schr\"{o}dinger equation ordered by Zhu-Kroemer. This ordering is known to describe abrupt heterojunctions in semiconductor materials.
\end{abstract}

\maketitle
\thispagestyle{empty}
\newpage
\section{Introduction}

In quantum mechanics, the Schr\"{o}dinger equation was initially treated considering physical systems with constant effective mass \cite{Schr,Griffiths}. Theoretical advances in solid-state physics have unleashed a new problem, i. e., the existence of non-relativistic particles that behave like a position-dependent effective mass \cite{Roos,Bastard,Weisbuch}. Later, the discussion of the concept of position-dependent mass (PDM) in a relativistic scenario was accomplished in Refs. \cite{Almeida,Almeida1,Almeida2}. The fact is that the concept of position-dependent effective mass has attracted the interest of many researchers, this is due to its applications, see f. e. Refs. \cite{Dong,Dong1,Navarro,Pourali,Kasapoglu}.

A feature of the PDM concept is that it becomes ambiguous in the non-relativistic quantum regime. Nonetheless, they are very important to describe some problems, such as, the study of electronic properties of semiconductor heterostructures \cite{Bastard}, harmonic oscillators \cite{Pourali}, impurities in crystals \cite{Luttinger}, some applications related to hermiticity from the Hamiltonian operator \cite{Mustafa}, to atomic and molecular physics \cite{Sever}, to supersymmetry \cite{Plastino}, particle dynamics in crystal structures \cite{Nabu}, among others.

The ambiguity of the kinetic energy operator (KEO) in non-relativistic models with PDM was discussed in Ref. \cite{Almeida}. The fact is that there is no agreement on the most appropriate way to order the KEO. Despite several literary discussions, several researchers and supporters of the concept of PDM propose that the most appropriate way to organize the KEO is
\begin{equation}
\hat{T}=-\frac{\hbar^2}{4}[m(\vec{r})^{\alpha}\vec{\nabla}m(\vec{r})^{\beta}\vec{\nabla}m(\vec{r})^{\gamma}+m(\vec{r})^{\gamma}\vec{\nabla}m(\vec{r})^{\alpha}\vec{\nabla}m(\vec{r})^{\beta}],
\end{equation}
where $\alpha$, $\beta$, and $\gamma$ are the parameters of von Roos (1983) \cite{Roos}, such that $\alpha+\beta+\gamma=-1$. The four most popular proposals in the literature are: BenDaniel-Duke (1966) \cite{BenDaniel}, Gora-Willian (1969) \cite{GoraW}, Zhu-Kroemer (1983) \cite{ZhuKroemer} and Li-Kuhn (1993) \cite{LiKuhn}.

Together with the advance of the concept of PDM, arise the theory of communication (or information theory), proposed by Claude E. Shannon in 1948 \cite{Shannon}. The construction of the concept of Shannon's entropy arises inside of a framework of the mathematical theory of communication \cite{Kripp}. This concept can be considered a section of probability theory that studies communication systems \cite{Zhou}, cryptography \cite{Grasselli}, and noise theory \cite{Amigo}. In the quantum scenario, the interpretation of the Shannon entropy in the position space $S_{x}$ is related to measuring the uncertainty of the particle location at position space \cite{Lima,Lima1}. Meanwhile, the Shannon entropy at reciprocal space $S_{p}$ is related to the uncertainty measurement of the particle's momentum. In this way, the Shannon entropy \cite{Shannon} is presented in quantum mechanics as a new formalism for the uncertainty measures of the particle's positions and momentum.

Another interesting element from communication theory and a precursor of the Shannon entropy is the Fisher information (or Fisher entropy) \cite{Fisher}. In the quantum context, Fisher information is intrinsically related to the uncertainty of a measurement, see Refs. \cite{Shi,Arenas,Ikot}. In other words, Fisher's information is a way of measuring the amount of information that certain observable carries concerning a parameter with an intrinsic probability related \cite{Fisher}.

In this work, we are interested in the calculation of quantum information entropies for a position-dependent mass (solitonic mass) and subject to a deformable hyperbolic potential (hyperbolic well) with KEO of  Zhu-Kroemer. It is important to mention that some studies have already been carried out using this mass profile and with the kinetic energy operator proposed by BenDaniel and Duke, see Ref. \cite{Cunha}. Throughout the work, the influence of the mass distribution in the system is discussed, the entropic quantities and the measurement uncertainties of the model are studied.

The work is organized as follows: In Sec. II, the one-dimensional PDM problem ordered by Zhu-Kroemer is discussed. In Sec III, some basic concepts about Shannon entropy are presented. Afterward, the calculate of the Shannon entropy is performed. In Sec. IV, the concept and results of the Fisher information of the model are demonstrated. Finally, in Sec. V we discuss our findings.


\section{A Schr\"{o}dinger problem with position-dependent mass}

Throughout this section, we study the analytical solutions of a solitonic mass distribution, i. e., a position-dependent mass with a soliton-like mass profile, namely,
\begin{equation}\label{mass}
m(x)=m_{0}sech^{2}(ax),
\end{equation}
where $m_{0}$ is the asymptotic mass of the system when $x\rightarrow 0$ and $\textit{a}$ is a parameter that adjusts the dimension of the hyperbolic secant argument and controls the width of the mass distribution. 

\begin{figure}[ht!]
\centering
\includegraphics[height=6cm,width=6.5cm]{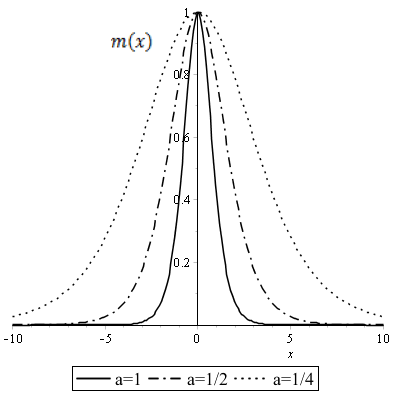}
\caption{Solitonic mass distribution for several values of parameter $a$ and $m_{0}$ constant.}
\label{m}
\end{figure}

Here it is interesting to mention that the solitons are structures that have drawn attention in several areas of physics from solid-state physics \cite{Heeger,Kartashov} to field theories \cite{Rajaraman}. These structures arise in a nonlinear theory, keeps their shape unchanged when interacts with another soliton, and have finite energy. The soliton mass distribution has the form of eq. (\ref{mass}) (for more details about the solitons, see Ref. \cite{Rajaraman}). The plot representation of the mass distribution investigated is shown in Fig. \ref{m}.

\begin{figure}[h]
\centering
\includegraphics[height=6cm,width=6.5cm]{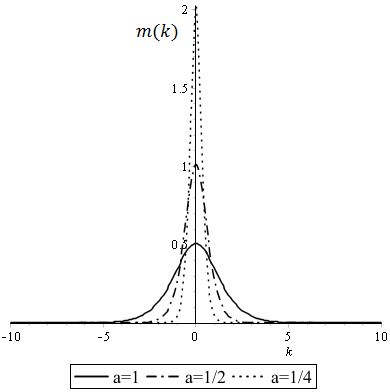}
\caption{Mass distribution at reciprocal space with $m_{0}$ constant.}
\label{m1}
\end{figure}

The mass distribution $m(x)$ in the reciprocal space is 
\begin{equation}
m(k)=\sqrt{\frac{\pi}{2}}\frac{m_{0}k}{a^{2}}csch\bigg(\frac{k\pi}{2a}\bigg),
\end{equation}
this expression gives us the mass distribution shown in Fig. \ref{m1}.

The energy of dispersion of this mass distribution is given by
\begin{equation}
E(k)=\sqrt{\frac{2}{\pi}}\frac{a^{2}\hbar^{2}}{m_{0}}\bigg[{k^{2}} _{2}F_{1}\bigg(\frac{1}{2}, \frac{3}{2};  \frac{3}{2}; \frac{k^{2}}{4} \bigg)-\cosh(k)\bigg].
\end{equation}

To study the quantum-mechanical system with PDM in a hyperbolic well, we consider the potential as follows:
\begin{equation}\label{eq1}
V(x)=V_{0}\sinh^{2}(x)+V_{1}\cosh^{2}(x),
\end{equation}
where the expression (\ref{eq1}) represents a confining potential, to appropriate values of $V_{0}$ and $V_{1}$. For convenience, let us restrict our study to the case of the positive values of $V_0$ and $V_1$, i. e., the case of a hyperbolic quantum well. Depending on the values of these parameters, the well transforms into a barrier.

Constructing a theory that the Hamiltonian is Hermitian, and considering that mass distribution describes an abrupt heterojunction of a material, as discussed in Ref. \cite{ZhuKroemer}, let us order the KEO of Schr\"{o}dinger's theory as
\begin{equation}
\hat{T}=-\frac{\hbar^{2}}{2}\frac{1}{m(x)^{1/2}}\frac{d^{2}}{dx^{2}}\frac{1}{m(x)^{1/2}},
\end{equation}
this operador was proposed by Zhu-Kroemer \cite{ZhuKroemer}.

Therefore, Schr\"{o}dinger's equation is
\begin{equation}\label{eq2}
-\frac{\hbar^{2}}{2m(x)}\frac{d^{2}\psi(x)}{dx^{2}}+\frac{\hbar^{2}}{2}\frac{m^{'}(x)}{m(x)^{2}}\frac{d\psi(x)}{dx}+\bigg[\frac{\hbar^{2}}{4}\frac{m^{''}(x)}{m(x)^{2}}-\frac{3 \hbar^{2}}{8}\frac{m^{'}(x)^{2}}{m(x)^{3}}+V(x)\bigg]\psi(x)=E\psi(x).
\end{equation}

The expression (\ref{eq2}) was proposed by Zhu and Kroemer \cite{ZhuKroemer} to describe a mass distribution in an abrupt heterojunction. 

It is essential to remember that in an abrupt junction of two materials, a particle restricted to the periodic potential, the wave function must respect the condition
\begin{align}\label{BD}
    \frac{1}{m_{L}^{*}}\frac{\partial\psi}{\partial x}\bigg\vert_{x\to x_{L}^{+}}=\frac{1}{m_{R}^{*}}\frac{\partial\psi}{\partial x}\bigg\vert_{x\to x_{R}^{+}},
\end{align}
where $x_j$ is the position of the well (or your wall). The subscripts $L$ and $R$ respectively indicate the left and right limits having the location of the wall, see Refs. \cite{BenDaniel,SK}. In fact, Eq. (\ref{BD}) is known as the BenDaniel-Duke boundary condition. In the case of an abrupt heterojunction described by a Hamiltonian with Zhu-Kromer ordering (Eq. (\ref{eq2})), a boundary condition analogous to the BenDaniel-Duke problem is
\begin{align}\label{CZK}
    \frac{1}{\sqrt{m_{L}^{*}}}\frac{\partial\psi}{\partial x}\bigg\vert_{x\to x_{L}^{+}}=\frac{1}{\sqrt{m_{R}^{*}}}\frac{\partial\psi}{\partial x}\bigg\vert_{x\to x_{R}^{+}}.
\end{align}
This condition of the derivatives of the wave function in a system with effective mass distribution guarantees the conservation of the current density of our \cite{Harrison,ZK1} system. Here, we advance that the solutions discussed in the future respect this boundary condition (\ref{CZK}). The case (\ref{CZK}) is valid for every case in which a bond chain produces bands described by the theory of effective mass at low energies. In truth, this condition can be appropriate to construct the wave functions of tight-binding through the junction between regions with effective mass differences.

\subsection{From the ordered equation to the ordinary Schr\"{o}dinger equation}

Allow us to make a small connection between the Zhu-Kroemer ordered model and the ordinary Schr\"{o}dinger theory. Let us start by considering the operator kinetic energy, i. e.,
\begin{equation}
\hat{T}=-\frac{\hbar^{2}}{2}\frac{1}{m(x)^{1/2}}\frac{d^{2}}{dx^{2}}\frac{1}{m(x)^{1/2}},
\end{equation}
where the Hamiltonian operator for stationary systems can be written as
\begin{equation}
\hat{H}=-\frac{\hbar^{2}}{2}\frac{1}{m(x)^{1/2}}\frac{d^{2}}{dx^{2}}\frac{1}{m(x)^{1/2}}+V(x).
\end{equation}

Therefore, the stationary Schr\"{o}dinger equation (\ref{eq2}) for a position-dependent mass ordered by Zhu-Kroemer is 
\begin{equation}
-\frac{\hbar^{2}}{2}\frac{1}{m(x)^{1/2}}\frac{d^{2}}{dx^{2}}\frac{1}{m(x)^{1/2}}\psi(x)+V(x)\psi(x)=E\psi(x),
\end{equation}
i. e.,
\begin{equation}\label{eee}
-\frac{\hbar^{2}}{2m(x)}\frac{d^{2}\psi(x)}{dx^{2}}+\frac{\hbar^{2}}{2}\frac{m^{'}(x)}{m(x)^{2}}\frac{d\psi(x)}{dx}+\bigg[\frac{\hbar^{2}}{4}\frac{m^{''}(x)}{m(x)^{2}}-\frac{3\hbar^{2}}{8}\frac{m^{'}(x)^{2}}{m(x)^{3}}+V(x)\bigg]\psi(x)=E\psi(x).
\end{equation}

For convenience, the variable change $\psi(x)=\sqrt{m(x)}\Theta(x)$ is considered in Eq. (\ref{eee}). In this way, Eq. (\ref{eee}) is rewritten as
\begin{equation}\label{eq3}
-\frac{\hbar^{2}}{2m(x)}\frac{d^{2}\Theta(x)}{dx^{2}}+V(x)\Theta(x)=E\Theta(x),
\end{equation}
where the equation (\ref{eq3}) is similar to Schr\"{o}dinger equation for constant mass. Therefore, here we can conclude that the effective potential due to the spatial distribution of mass is equivalent to the ordinary potential of the system in the space of $\Theta(x)$. In truth, this is the great advantage of using the Zhu-Kroemer ordering \cite{ZhuKroemer}. Despite the BenDaniel-Duke ordering \cite{BenDaniel} being more used, it produces an effective potential due to the mass distribution making the problem more complex. Meanwhile, the effective potential of the Zhu-Kroemer model in the transformed space (i. e., the Schr\"{o}dinger equation with PDM for $\Theta$) is the trivial case of the Schr\"{o}dinger equation with position-dependent mass.

\subsection{The hyperbolic well}

For the solitonic mass distribution (\ref{mass}) and the hyperbolic potential (\ref{eq1}) with the positive parameters $V_{0}$ and $V_{1}$, we have that the equation that describes our theory is
\begin{equation}
\frac{1}{sech^{2}(x)}\frac{d^{2}\Theta(x)}{dx^{2}}+[\delta-\alpha\sinh^{2}(x)-\beta\cosh^{2}(x)]\Theta(x)=0,
\end{equation}
i. e., 
\begin{equation}\label{eq4}
\frac{d^{2}\Theta(x)}{dx^{2}}+[\delta \\ \ sech^{2}(x)- \alpha\tanh^{2}(x)-\beta]\Theta(x)=0,
\end{equation}
where
\begin{align} 
\delta=\frac{2m_{0}E}{a^{2}\hbar^{2}}; \hspace{1cm} \alpha=\frac{2m_{0}V_{0}}{a^{2}\hbar^{2}}; \hspace{1cm} \beta=\frac{2m_{0}V_{1}}{a^{2}\hbar^{2}}.
\end{align}
In fact, perceive that the Schr\"{o}dinger equation ordered by Zhu and Kroemer \cite{ZhuKroemer} with mass the solitonic mass distribution $m(x)$ leads to an eigenvalue problem with a Poschl-Teller-like potential, see Ref. \cite{PT}.

\begin{figure}[ht!]
\centering
\includegraphics[height=6cm,width=6.5cm]{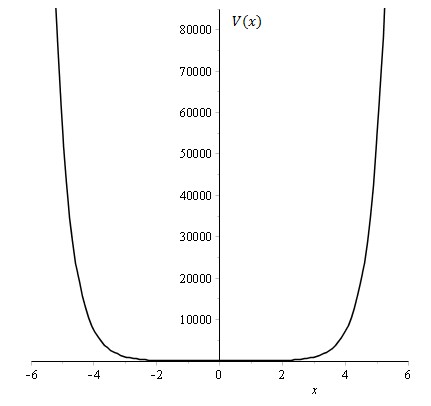}
\caption{Plot of potential $V(x)=V_{0}\sinh^{2}(x)+V_{1}\cosh^{2}(x)$ with $V_{0}$ and $V_{1}$ positive.}\label{fig_pot}
\end{figure}

The simplification of Eq. (\ref{eq4}) leads us to
\begin{equation}
\frac{d^{2}\Theta(x)}{dx^{2}}+[(\delta+\alpha) \\ \ sech^{2}(x)- (\alpha+\beta)]\Theta(x)=0.
\end{equation}

Here it is convenient to assume the variable change, namely,
\begin{align} 
z\rightarrow\tanh(x), \hspace{1cm} \text{with} \hspace{1cm} -1 \leq z \leq 1,
\end{align}
so that
\begin{equation}\label{eq6}
(1-z^{2})\frac{d^{2}\Theta(z)}{dz^{2}}-2z\frac{d\Theta(z)}{dz}+\bigg[\delta+\alpha-\bigg(\frac{\alpha+\beta}{1-z^{2}}\bigg)\bigg]\Theta(z)=0.
\end{equation}

Equations that are written in the form:
\begin{equation}\label{eq5}
(1-y^{2})\frac{d^{2}L(y)}{dy^{2}}-2y\frac{dL(y)}{dy}+\bigg[s(s+1)-\bigg(\frac{\kappa^{2}}{1-y^{2}}\bigg)\bigg]L(y)=0,
\end{equation}
are known as the associated Legendre equation. Its solutions are the associated Legendre functions, i. e.,
\begin{equation}
L^{(1)}(y)=\frac{1}{\Gamma(1-\kappa)}\bigg[\frac{1+y}{1-y}\bigg]^{\kappa/2} \\ \ _{2}F_{1}\bigg(-s,s+1;1-\kappa;\frac{1-y}{2}\bigg),
\end{equation}
\begin{equation}
L^{(2)}(y)=\frac{\sqrt{\pi}}{2^{s+1}}\frac{\Gamma(s+\kappa+1)}{\Gamma(s+3/2)}\frac{(1-y^{2})^{\kappa/2}}{y^{s+\kappa+1}} \\ \ _{2}F_{1}\bigg(\frac{s+\kappa+1}{2}, \frac{s+\kappa+2}{2}; s+\frac{3}{2}; \frac{1}{y^{2}}\bigg).
\end{equation}

Comparing to Eq. (\ref{eq5}) with (\ref{eq6}) it is observed that
\begin{align}
s=\frac{1}{2}-\frac{1}{2}\sqrt{1+4(\delta+\alpha)}, \hspace{0.5cm} \text{and} \hspace{0.5cm} \kappa^{2}=\alpha+\beta.
\end{align}

Therefore, the stationary solutions of the model are given by
\begin{align}\label{s1}
&\Theta^{(1)}(z)=\frac{1}{\Gamma(1-\kappa)}\bigg[\frac{1+z}{1-z}\bigg]^{\kappa/2}\hspace{0.1cm} _{2}F_{1}\bigg(-s,s+1;1-\kappa;\frac{1-z}{2}\bigg),\\
&\Theta^{(2)}(z)=\frac{\sqrt{\pi}}{2^{s+1}}\frac{\Gamma(s+\kappa+1)}{\Gamma(s+3/2)}\frac{(1-z^{2})^{\kappa/2}}{z^{s+\kappa+1}}\hspace{0.1cm}_{2}F_{1}\bigg(\frac{s+\kappa+1}{2}, \frac{s+\kappa+2}{2}; s+\frac{3}{2}; \frac{1}{z^{2}}\bigg).
\end{align}

By rewriting the solutions in terms of $\psi(x)$, we have that
\begin{align}
\psi^{(1)}(x)=&A\frac{sech(ax)}{\Gamma(1-\kappa)}\bigg[\frac{1+\tanh(ax)}{1-\tanh(ax)}\bigg]^{\kappa/2}\hspace{0.1cm} _{2}F_{1}\bigg(-s,s+1;1-\kappa;\frac{1-\tanh(ax)}{2}\bigg),\\ \nonumber
\psi^{(2)}(x)=&B\frac{\sqrt{\pi}}{2^{s+1}}\frac{\Gamma(s+\kappa+1)}{\Gamma(s+3/2)}\frac{(1-\tanh^{2}(ax))^{\kappa/2}}{\tanh(x)^{s+\kappa+1}}\hspace{0.1cm}\times \\
&_{2}F_{1}\bigg(\frac{s+\kappa+1}{2}, \frac{s+\kappa+2}{2}; s+\frac{3}{2}; \frac{1}{\tanh^{2}(ax)}\bigg),
\end{align}
with $A$ and $B$ system normalization constants.

For wave functions to be physically acceptable, normalization of the eigenfunctions is required. To preserve the normalization, $\psi^{(2)}(x)$ is disregarded, as this function tends to diverge at the origin of the system. Similarly, for that te wave function $\psi^{(1)}(x)$ conserve the system normalization, it is required that
\begin{equation}\label{eq8}
\lim_{x\rightarrow\pm\infty} \\ \ _{2}F_{1}\bigg(-s,s+1;1-\kappa;\frac{1-\tanh(ax)}{2}\bigg)=0,
\end{equation}
for the previous condition to be preserved, it is necessary that
\begin{align}\label{eq9}
s=n, \hspace{1cm} \text{with} \hspace{1cm} n=0,1,2,...
\end{align}
and
\begin{align}
\vert\kappa\vert \leq n,
\end{align}
with $\kappa$ interger.

The Eq. (\ref{eq9}) leads directly to the quantization of energy in the system, i. e.,
\begin{align}\label{energia}
E_{n}=\frac{a^{2}\hbar^{2}}{2m_{0}}n(n+1)-V_{0}, \hspace{1cm} \text{with} \hspace{1cm} n=0,1,2,3,...
\end{align}
here it is interesting to mention that the model has $2n+1$ degenerate states.

Finally, we observe that the symmetric and antisymmetric solutions for the wave function are described by
\begin{equation}
\psi_{n}^{\kappa}(x)= A_{n}^{\kappa}\frac{sech(ax)}{\Gamma(1-\kappa)}\bigg[\frac{1+\tanh^{2}(ax)}{1-\tanh^{2}(ax)}\bigg]^{\kappa/2} \\ \ _{2}F_{1}\bigg(-n,n+1;1-\kappa;\frac{1-\tanh(ax)}{2}\bigg),
\end{equation}
with $A_{n}^{\kappa}$ determined by normalization.

\begin{figure}[ht!]
\centering
\includegraphics[height=6cm,width=6.5cm]{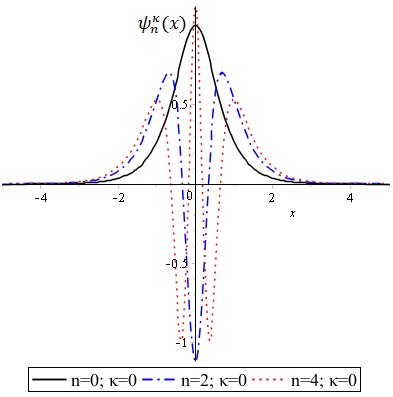}
\includegraphics[height=6cm,width=6.5cm]{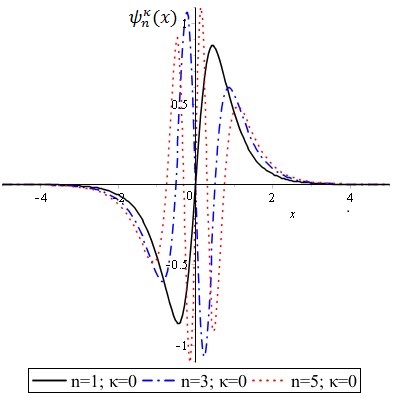}\\
(a) \hspace{6cm} (b)
\caption{(a) Symmetrical wave eigenfunctions. (b) Antisymmetric wave eigenfunctions.}\label{ffig}
\end{figure}

\begin{figure}[ht!]
\centering
\includegraphics[height=4cm,width=4.5cm]{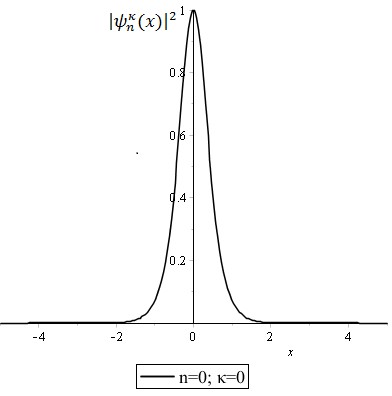}
\includegraphics[height=4cm,width=4.5cm]{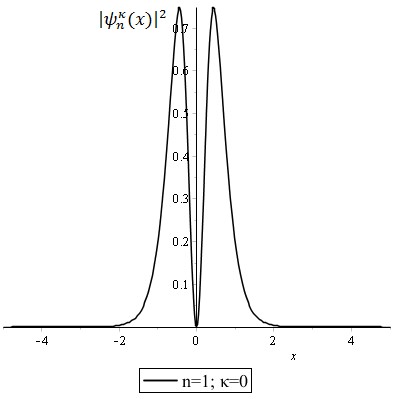}
\includegraphics[height=4cm,width=4.5cm]{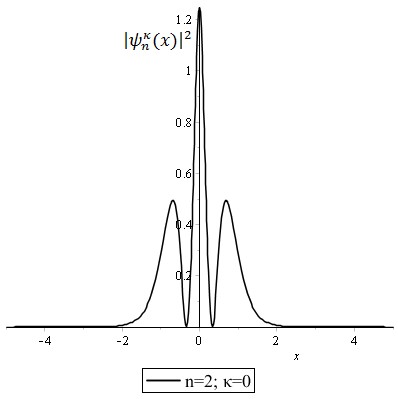}\\
\includegraphics[height=4cm,width=4.5cm]{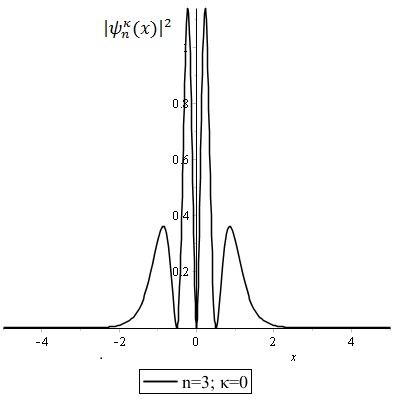}
\includegraphics[height=4cm,width=4.5cm]{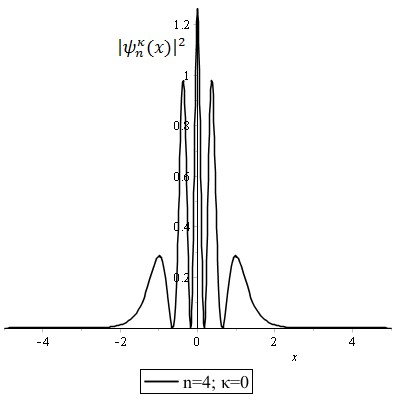}
\caption{Probability densities of solitonic mass distribution when subjected to hyperbolic potential.}\label{ffig1}
\end{figure}

The plots of the solutions for the symmetric and antisymmetric cases and their respective probability densities when $a=2$ and $\kappa=0$ are shown in Figs. \ref{ffig} and \ref{ffig1}.

\section{Shannon's entropy}

Based on thermodynamics, ``the concept of entropy provides a measure of the irreversibility of the physical system" \cite{Callen}. According to statistical mechanics, the concept of statistical entropy appears as a measure associated with the degree of disorder of the system \cite{Callen,Pathria}.

The concept of Shannon entropy was first proposed in 1948 by Claude E. Shannon, at work entitled, \textit{``A mathematical theory of communication"} \cite{Shannon}, where Shannon describes the entropy as an element of the information theory (or communication theory). To the eyes of information theory, entropy provides a measure of uncertainty in a given probability distribution. This quantity that describes the uncertainty measure is called Shannon's entropy.

Shannon's entropy can be interpreted as a starting point for measuring the uncertainty of a probability distribution associated with an information source, see f. e. Refs. \cite{Hirschmann,Beckner,Navarro}. According to Born \cite{Born}, the statistical interpretation of the quantum system is given by
\begin{equation}\label{eq}
\rho(x)dx=\vert\Psi(x,t)\vert^{2}dx=\vert\psi(x)\vert^{2}dx,
\end{equation}
where Eq. (\ref{eq}) describes the probability of finding the particle in the state $\Psi(x,t)$ between the spatial interval $x$ and $x+dx$ \cite{Born}. The quantity $\vert\Psi(x,t)\vert^{2}$ is the probability density of the system in $(1+1)D$ \cite{Cohen}. For a probability density of a continuous system in position space, the Shannon entropy for the $n$-th state is defined as
\begin{equation}\label{shan}
S_{x}^{n}=-\int_{-\infty}^{\infty}\vert\psi_{n}(x)\vert^{2}\text{ln}(\vert\psi_{n}(x)\vert^{2})dx.
\end{equation} 

Similarly, for the reciprocal space, the entropy is
\begin{equation}\label{shan1}
S_{p}^{n}=-\int_{-\infty}^{\infty}\vert\Phi_{n}(p)\vert^{2}\ln(\vert\Phi_{n}(p)\vert^{2})dp,
\end{equation}
where
\begin{equation}
\Phi_{n}(p)=\frac{1}{\sqrt{2\pi}}\int_{-\infty}^{\infty}\psi_{n}(x)\text{e}^{-ipx}dx.
\end{equation}

Shannon entropy plays an important role in the Heisenberg uncertainty measure \cite{Griffiths,Lima,Lima1}. The relation of entropic uncertainty related to position and momentum was initially obtained by Beckner (1975) \cite{Beckner}, Bialynicki-Birula and Mycielski (1975) \cite{Bialy}. The uncertainty relation is 
\begin{equation}
S_{x}^{n}+S_{p}^{n}\geq D(1+\ln\pi),
\end{equation}
where $D$ is the dimension of the spatial coordinates of the system.

\subsection{Shannon's entropy of the mass distribution at well}

To calculate the Shannon entropy in the quantum-mechanical system with position-dependent mass (\ref{mass}), and subject to the potential (\ref{eq1}), it is necessary to consider the solutions of the Schr\"{o}dinger equation ordered by Zhu-Kroemer. As discussed in Sec. II, the complete set of solutions are described by:
\begin{equation}\label{sol}
\psi_{n}^{\kappa}(x)=A_{n}^{\kappa}\frac{sech(ax)}{\Gamma(1-\kappa)}\bigg[\frac{1+\tanh^{2}(ax)}{1-\tanh^{2}(ax)}\bigg]^{\kappa/2} \\ \ _{2}F_{1}\bigg(-n, n+1, 1-\kappa; \frac{1-\tanh(ax)}{2}\bigg),
\end{equation}
where, 
\begin{align}
n=1,2,3,... \hspace{1.5cm} \kappa=-n,-n+1,...,-1,0,1,...,n-1,n.
\end{align} 

\begin{figure}[ht!]
\centering
\includegraphics[scale=0.6]{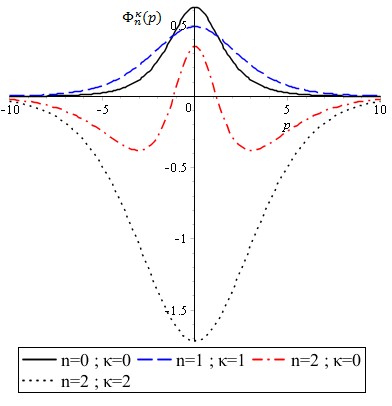}\\
\includegraphics[height=3.8cm,width=4.2cm]{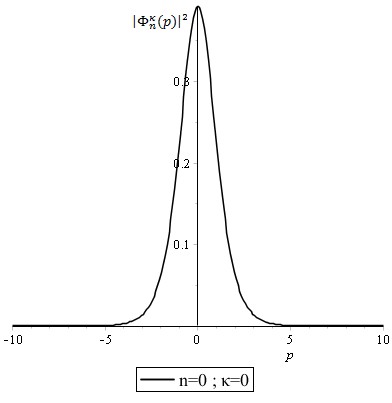}
\includegraphics[height=3.8cm,width=4.2cm]{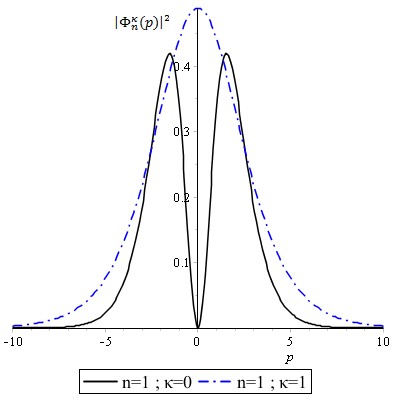}
\includegraphics[height=3.8cm,width=4.2cm]{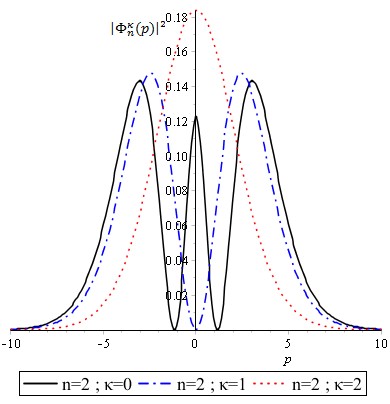}
\includegraphics[height=3.8cm,width=4.2cm]{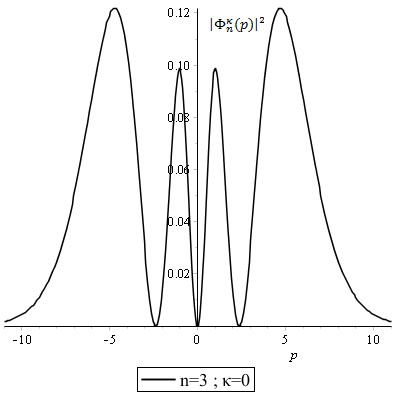}
\caption{Plots of eigenfunctions in reciprocal space for the first energy levels, i. e., $n=1,2,$ and $3$.}
\label{ffig2}
\end{figure}

Indeed, to calculate the Shannon Entropy, the Eq. (\ref{sol}) is expanded to the first energy levels, and the investigation of the eigenfunctions in the reciprocal space is performed. The eigenfunctions at position and momentum space are shown in table \ref{tab1}. The plots of the real eigenfunctions at momentum space and the probability densities of the system are exhibited in Fig. \ref{ffig2}.

\begin{table}[ht!] 
\centering
\caption{Eigenfunctions and energy eigenvalues of the model.}
\label{tab1}
\resizebox{12cm}{!}{%
\begin{tabular}{|c|c|c|c|c|}
\hline
$n$ & $\kappa$ & $\psi_{n}^{\kappa}(x)$ & $\Phi_{n}^{\kappa}(p)$ & $E_{n}$ \\ \hline
$0$ & $0$ & $\sqrt{\frac{a}{2}}\sech(ax)$ & $\sqrt{\frac{\pi}{4a}}\sech(\frac{p\pi}{2a})$ & $-V_{0}$ \\ \hline
\multirow{2}{*}{1} & $0$ & $\sqrt{\frac{3a}{2}}\sech(ax)\tanh(ax)$ & $ip\sqrt{\frac{3\pi}{4a^{3}}} \sech(\frac{p\pi}{2a})$ & $\frac{a^{2}\hbar^{2}}{m_{0}}-V_{0}$ \\ \cline{2-5}
 & $1$ & $\sqrt{\frac{3a}{4}}\sech^{2}(ax)$ & $p\sqrt{\frac{3\pi}{8a^{3}}}\csch(\frac{p\pi}{2a})$ & $\frac{a^{2}\hbar^{2}}{m_{0}}-V_{0}$ \\ \hline
\multirow{3}{*}{2} & $0$ & $\sqrt{\frac{5a}{8}}[2\cosh^{2}(ax)-3]\sech^{3}(ax)$ & $\sqrt{\frac{5\pi}{64a^{5}}}(a^{2}-3p^{2})\sech(\frac{p\pi}{2a})
$ & $\frac{3a^{2}\hbar^{2}}{m_{0}}-V_{0}$ \\ \cline{2-5}
 & $1$ & $-\sqrt{\frac{15a}{4}}\sech^{2}(ax)\tanh(ax)$ & $-ip^{2}\sqrt{\frac{15\pi}{32a^{5}}}\csch(\frac{p\pi}{2a})$ & $\frac{3a^{2}\hbar^{2}}{m_{0}}-V_{0}$ \\ \cline{2-5}
 & $2$ & $-\sqrt{\frac{15a}{16}}\sech^{3}(ax)$ & $-\sqrt{\frac{15\pi}{128a^{5}}}(a^{2}+p^{2})\sech(\frac{p\pi}{2a})$ & $\frac{3a^{2}\hbar^{2}}{m_{0}}-V_{0}$ \\ \hline
$3$ & $0$ & $\sqrt{\frac{7a}{8}}[2\cosh^{2}(ax)-5]\tanh(ax)\sech^{3}(ax)$ & $ip\sqrt{\frac{7\pi}{576a^{7}}}(7a^{2}-5p^{2})\sech(\frac{p\pi}{2a})$ & $\frac{6a^{2}\hbar^{2}}{m_{0}}-V_{0}$ \\ \hline
\end{tabular}}
\end{table}

The numerical study of the Shannon entropy was carried out considering the eigenfunctions at position and momentum space, as well as the definition of the Shannon entropy, i. e., Eqs. (\ref{shan}) and (\ref{shan1}). The numerical result of the Shannon entropy is presented in the table \ref{tab2}.

\begin{table}[ht!]\centering
\caption{Numerical results of the Shannon entropy.}
\label{tab2}
\resizebox{12cm}{!}{%
\begin{tabular}{|c|c|c|c|c|c|c|}
\hline
$n$ & $\kappa$ & $a$ & $S_{x}$ & $S_{p}$ & $S_{x}+S_{p}$ & $1+\ln \pi$   \\ \hline
 &  & $2$ & $2-\ln(4)=0,6137$ & $2-\ln(\pi/2)=1,5484$ & $4-\ln(2\pi)=2,1622$ & $2,1447$ \\
$0$ & $0$ & $4$ & $2-\ln(8)=-0,0794$ & $2-\ln(\pi/4)=2,2415$ & $4-\ln(2\pi)=2,1622$ & $2,1447$ \\
 &  & $6$ & $2-\ln(12)=-0,4849$ & $2-\ln(\pi/6)=2,6470$ & $4-\ln(2\pi)=2,1622$ & $2,1447$ \\ \hline 
 \multirow{6}{*}{1} &  & $2$ & $10/3-\ln(12)=0,8484$ & $2,1088$ & $2,9572$ & $2,1447$ \\ 
  & $0$ & $4$ & $10/3-\ln(24)=0,1552$ & $2,8019$ & $2,9572$ & $2,1447$ \\
  &  & $6$ & $10/3-\ln(36)=-0,2501$ & $3,2074$ & $2,9572$ & $2,1447$ \\ \cline{2-7}
  &  & $2$ & $10/3-\ln(24)=0,1552$ & $1,9960$ & $2,1512$ & $2,1447$ \\
  & $1$ & $4$ & $10/3-\ln(48)=-0,5378$ & $2,6891$ & $2,1512$ & $2,1447$ \\
  &  & $6$ & $10/3-\ln(72)=-0,9433$ & $3,0946$ & $2,1512$ & $2,1447$ \\ \hline
 \multirow{9}{*}{2} &  & $2$ & $0,8571$ & $2,4707$ & $3,3278$ & $2,1447$ \\
  & $0$ & $4$ & $0,1640$ & $3,1638$ & $3,3278$ & $2,1447$ \\
  &  & $6$ & $-0,2415$ & $3,5693$ & $3,3278$ & $2,1447$ \\ \cline{2-7}
  &  & $2$ & $0,4125$ & $2,42469$ & $2,8372$ & $2,1447$ \\
  & $1$ & $4$ & $-0,2806$ & $3,1178$ & $2,8372$ & $2,1447$ \\
  &  & $6$ & $0,6861$ & $3,5233$ & $2,8372$ & $2,1447$ \\ \cline{2-7}
  &  & $2$ & $-0,0874$ & $2,2355$ & $2,1481$ & $2,1447$ \\
  & $2$ & $4$ & $-0,07806$ & $2,9287$ & $2,1481$ & $2,1447$ \\
  &  & $6$ & $-1,1861$ & $3,3342$ & $2,1481$ & $2,1447$ \\ \hline
 \multirow{3}{*}{3} &  & $2$ & $0,8555$ & $2,7265$ & $3,5820$ & $2,1447$ \\
  & $0$ & $4$ & $0,1625$ & $3,4195$ & $3,5820$ & $2,1447$ \\
  &  & $6$ & $-0,2428$ & $3,8248$ & $3,5820$ & $2,1447$ \\ \hline
\end{tabular}}
\end{table}

Analyzing the Shannon entropy results (see table \ref{tab2}), we observe that it tends to decrease in the position space as the width of the mass distribution increases. Meanwhile, Shannon entropy increases in momentum space as the width of the mass distribution increases. An interesting result arises in the model, i. e., the quantity $S_{x}+S_{p}$ is a constant independent of the width of the spatial distribution of mass (this occurs in all eigenstates of this system). Further physical interpretations of this result, will be discussed in the final remarks.


 \section{Fisher's Information}

Communication theory has been studied in several areas of research \cite{Nalewajski,Nagaoka,Wang,Lian,Falaye0}, including quantum mechanics \cite{Rothstein,Zou}. In the quantum regime, information theory seeks to describe and make use of the different possibilities for processing and communicating information in a quantum-mechanical system \cite{Falaye,Serrano}. The fact that the properties of quantum systems differ from classical objects offers an opportunity for new communication protocols.

The Fisher information concept was introduced in 1925 \cite{Fisher}, its basic properties are not yet fully known. Despite its early origins in 1925, the most relevant contributions of Fisher's information theory emerged around 1980 \cite{Sears}. As discussed in Ref. \cite{Frieden},  it is interesting to note that the KEO in quantum mechanics can be considered as a measure of the distribution of information.

Fisher's information has reached a prominent position in several areas, for example, using Fisher's minimum information principle, is obtained non-relativistic quantum mechanics equations \cite{Frieden1,Reginatto}, i. e., the time-independent Kohn-Sham, and the time-dependent Euler equations \cite{Naje}. One application that has drawn the attention of researchers in the study of the Fisher information for systems with PDM \cite{Falaye}, is due to the large applications of the concept of  PDM in quantum mechanics, see f. e. Refs. \cite{Almeida,Almeida1,Almeida2}.

Fisher's information for an $\xi$ observable is defined (see Ref. \cite{Falaye}) as
\begin{equation}\label{Fisher}
F_{\xi}=\int_{-\infty}^{\infty}\rho(\xi)\bigg[\frac{d}{d\xi}\ln\vert\rho(\xi)\vert\bigg]^2d\xi=\int_{-\infty}^{\infty}\frac{[\rho^{'}(\xi)]^2}{\rho(\xi)}d\xi>0.
\end{equation}

To calculate Fisher's information for an observable $\xi$ with probability density $\vert\Psi(\xi,t)\vert^2$, we have
\begin{equation}
F_{\xi}=\int_{-\infty}^{\infty}\vert\Psi(\xi,t)\vert^2\bigg[\frac{d}{d\xi}\ln(\vert\Psi(\xi,t)\vert^2)\bigg]^{2} d\xi > 0,
\end{equation}
where for one-dimensional stationary quantum systems, at position space, $\rho(\xi)=\vert\Psi(\xi,t)\vert^2\simeq \vert\psi(\xi)\vert^2$ is the probability density. 

Meanwhile, Fisher's information at reciprocal space is given by
\begin{equation}
F_{p}^{n}=\int_{-\infty}^{\infty}\vert\Psi(p,t)\vert^2\bigg[\frac{d}{dp}\ln(\vert\Psi(p,t)\vert^2)\bigg]^{2} dp>0.
\end{equation} 

\subsection{Fisher's information of the mass distribution at well}

To perform the Fisher information calculation, let us consider again the eigenfunctions (37) (or the eigenfunctions of the first energy eigenstates of the model, see Table I) and rewrite the definition of the Fisher information (39) at both spaces, as follows:
\begin{align}\label{F1}
&F_{x}^{n}=4\int_{-\infty}^{\infty}\psi_{n}(x)\psi_{n}^{*'}(x) dx+\int_{-\infty}^{\infty}\bigg[\frac{\psi_{n}^{'}(x)}{\psi_{n}(x)}-\frac{\psi_{n}^{*'}(x)}{\psi_{n}^{*}(x)}\bigg]\vert\psi(x)\vert^{2}dx>0,\\ \label{F2}
&F_{p}^{n}=4\int_{-\infty}^{\infty}\Phi_{n}(p)\Phi_{n}^{*'}(p) dp+\int_{-\infty}^{\infty}\bigg[\frac{\Phi_{n}^{'}(p)}{\Phi_{n}(p)}-\frac{\Phi_{n}^{*'}(p)}{\Phi_{n}^{*}(p)}\bigg]\vert\Phi(p)\vert^{2}dp>0.
\end{align}

Let us also remember that the standard deviation of the position and momentum measures is given respectively by, 
\begin{align}\label{v}
\sigma_{x}^{2}=\langle x^{2}\rangle-\langle x\rangle^{2}, \hspace{1.5cm}
\sigma_{p}^{2}=\langle p^{2}\rangle-\langle p\rangle^{2},
\end{align}
where $\langle x\rangle$, $\langle x^{2}\rangle$, $\langle p\rangle$, and $\langle p^{2}\rangle$ are the expected values of $x$, $ x^{2}$, $p$ and $p^{2}$, see Refs. \cite{Cohen,Griffiths}. 

Using the definitions of Fisher's information presented in Eqs. (\ref{F1}) and (\ref{F2}) and the definitions of the standard deviation shown in Eq. (\ref{v}), the result of the relations uncertainty and fisher information measure are obtained and presented in Table \ref{tab3}.

\begin{table}[ht!]
\centering
\caption{Results for the uncertainty relation and Fisher information measure.}
\label{tab3}
\resizebox{10cm}{!}{%
\begin{tabular}{|c|c|c|c|c|c|c|}
\hline 
$n$ & $\kappa$ & $a$ & $F_{x}^{n\kappa}$ & $F_{p}^{n\kappa}$ & $\sigma_{x}^{2}$ & $\sigma_{p}^{2}$ \\ \hline
\multirow{3}{*}{0} & \multirow{3}{*}{0} & $2$ & $16/3$ & $\pi^{2}/12$ & $\pi^{2}/48$ & $4/3$ \\
 &  & $4$ & $64/3$ & $\pi^{2}/48$ & $\pi^{2}/192$ & $16/3$ \\
 &  & $6$ & $48$ & $\pi^{2}/108$ & $\pi^{2}/432$ & $36/3$ \\ \hline 
\multirow{6}{*}{1} & \multirow{3}{*}{0} & $2$ & $112/5$ & $(12+\pi^{2})/12$ & $(12+\pi^{2})/48$ & $28/5$ \\
 &  & $4$ & $448/5$ & $(12+\pi^{2})/48$ & $(12+\pi^{2})/196$ & $112/5$ \\
 &  & $6$ & $1008/5$ & $(12+\pi^{2})/108$ & $(12+\pi^{2})/432$ & $252/5$ \\ \cline{2-7}
 & \multirow{3}{*}{1} & $2$ & $64/5$ & $(\pi^{2}-6)/12$ & $(\pi^{2}-6)/48$ & $16/5$ \\
 &  & $4$ & $256/5$ & $(\pi^{2}-6)/48$ & $(\pi^{2}-6)/192$ & $64/5$ \\
 &  & $6$ & $576/5$ & $(\pi^{2}-6)/108$ & $(\pi^{2}-6)/432$ & $144/5$ \\ \hline
\multirow{9}{*}{2} & \multirow{3}{*}{0} & $2$ & $1136/21$ & $(15+\pi^{2})/12$ & $(15+\pi^{2})/48$ & $284/21$ \\
 &  & $4$ & $4544/21$ & $(15+\pi^{2})/48$ & $(15+\pi^{2})/192$ & $1136/21$ \\
 &  & $6$ & $3408/7$ & $(15+\pi^{2})/108$ & $(15+\pi^{2})/432$ & $852/7$ \\ \cline{2-7}
 & \multirow{3}{*}{1} & $2$ & $320/7$ & $\pi^{2}/12$ & $\pi^{2}/48$ & $80/7$ \\
 &  & $4$ & $1280/7$ & $\pi^{2}/48$ & $\pi^{2}/192$ & $320/7$ \\
 &  & $6$ & $2880/7$ & $\pi^{2}/108$ & $\pi^{2}/432$ & $720/7$ \\ \cline{2-7}
 & \multirow{3}{*}{2} & $2$ & $144/7$ & $(2\pi^{2}-15)/24$ & $(2\pi^{2}-15)/96$ & $36/7$ \\
 &  & $4$ & $576/7$ & $(2\pi^{2}-15)/96$ & $(2\pi^{2}-15)/384$ & $144/7$ \\
 &  & $6$ & $1296/7$ & $(2\pi^{2}-15)/216$ & $(2\pi^{2}-15)/864$ & $324/7$ \\ \hline
\multirow{3}{*}{3} & \multirow{3}{*}{0} & $2$ & $4592/45$ & $(49+3\pi^{2})/36$ & $(49+3\pi^{2})/144$ & $1148/45$ \\ 
 &  & $4$ & $18368/45$ & $(49+3\pi^{2})/144$ & $(49+3\pi^{2})/576$ & $4592/45$ \\
 &  & $6$ & $4592/5$ & $(49+3\pi^{2})/324$ & $(49+3\pi^{2})/1296$ & $1148/5$ \\ \hline
\end{tabular}}
\end{table}

With the obtained results, we note the existence of an ``information propagation" for the solitonic mass distribution when subjected to the confining potential $V(x)=V_{0}\sinh^{2}(x)+V_ {1}\cosh^{2}(x)$. We observe that the Fisher information tends to increase with the width of the mass distribution, i. e., at position space the information increases proportionally to the term of $a^{2}$. On the other hand, the information tends to decrease with the width of the mass distribution, i. e., it decreases with a factor proportional to the $a^{-2}$ at momentum space. Finally, with the results of standard deviation, the Heisenberg uncertainty principle was investigated, as a consequence, the following relations were obtained
\begin{align}
F_{x}=4\sigma_{p}^{2}, \hspace{0.5cm} \text{and} \hspace{0.5cm}
F_{p}=4\sigma_{x}^{2}.
\end{align}

In this way, the Heisenberg uncertainty principle is written as
\begin{equation}
F_{x}F_{p}\geq4\hbar^{2}.
\end{equation} 

\begin{figure}[ht!]
\centering
\includegraphics[height=4.5cm,width=6cm]{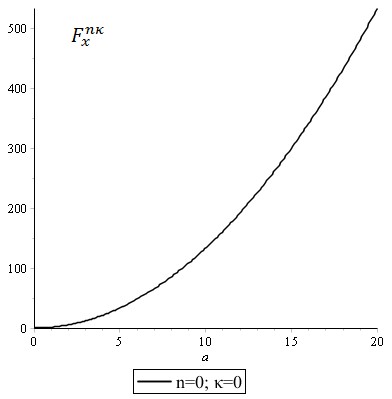}
\includegraphics[height=4.5cm,width=6cm]{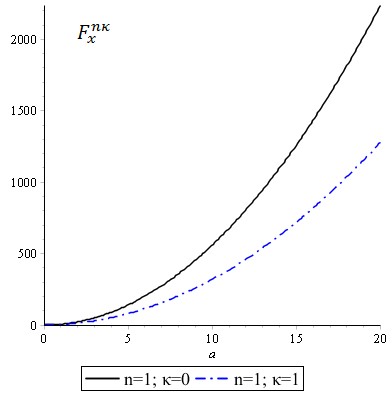}\\
\includegraphics[height=4.5cm,width=6cm]{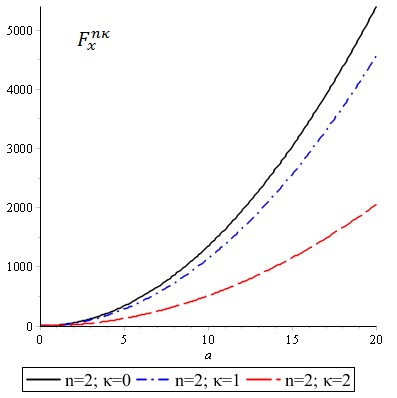}
\includegraphics[height=4.5cm,width=6cm]{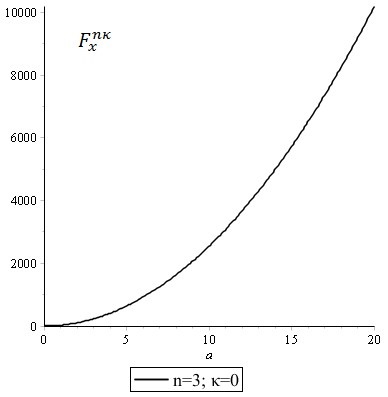}
\caption{Plots of the Fisher information as function of the width of the mass distribution at position space.}
\label{fff}
\end{figure}

\begin{figure}[ht!]
\centering
\includegraphics[height=4.5cm,width=6cm]{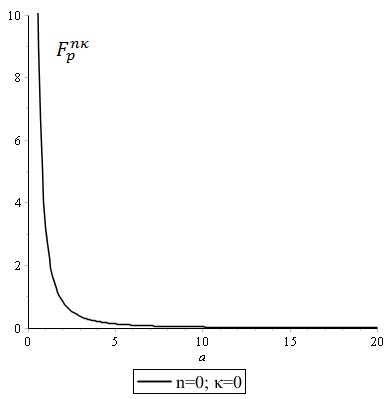}
\includegraphics[height=4.5cm,width=6cm]{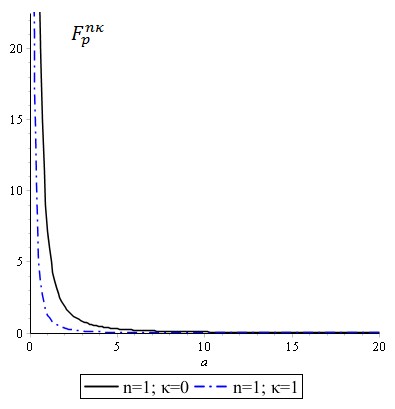}\\
\includegraphics[height=4.5cm,width=6cm]{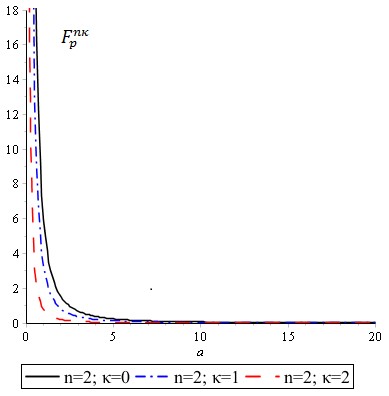}
\includegraphics[height=4.5cm,width=6cm]{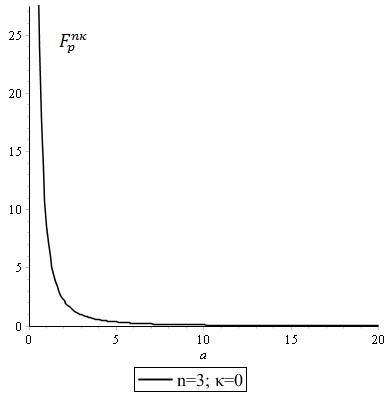}
\caption{Plots of the Fisher information as function of the width of the mass distribution at momentum space.}
\label{fff1}
\end{figure}

Finally, the plots of Fisher information as a function of the width of the mass distribution are shown in Figs. \ref{fff} and \ref{fff1}.

\section{Final remarks}

In this work, Shannon entropy and Fisher information for a solitonic mass distribution were investigated. The analytical solutions of the stationary Schr\"{o}dinger equation with position-dependent mass when subjected to the hyperbolic potential $V(x)=V_{0}\sinh^{2}(x)+V_{1}\cosh^{2}(x)$ (with $V_0$ and $V_1$ positive) were investigated. It was shown that the wave eigenfunctions for the solitonic mass ordered by Zhu-Kroemer is described by the associated Legendre functions.

Once the analytical solutions of the model are known, the Shannon entropy for the first energy levels was studied. We observe in the numerical results that the Shannon entropy tends to decrease exponentially with the width of the mass distribution at position space. On the other hand, the Shannon entropy at momentum space tends to increase proportionally, making the sum $S_{x}+S_{p}$ constant (for each state) concerning the width of the distribution, i. e.,
\begin{align}\label{222}
\frac{d}{da}(S_{x}+S_{p})=0.
\end{align}

In other words, the expected value of the information contained in each wave packet tends to increase significantly as the solitonic mass distribution tends to become more localized.

This result is a consequence of the symmetry of the mass distribution used. In fact, the amount $S_x+S_p$ becomes constant because when the parameter $a$ increases, i. e. the width of the mass distribution increases, the uncertainty in position measurements increases. However, in the reciprocal space, when the parameter $a$ increases, the effective mass becomes more localized, decreasing the uncertainties of measurements of the momentum. Furthermore, for each state, the Heisenberg uncertainty principle is constant. Therefore, the constancy of the amount $S_x+S_p$ in Eq. (\ref{222}) is a consequence of the Heisenberg uncertainty principle.

By investigating Fisher's information, we observe that the information tends to increase proportionally to the factor $a^{2}$ (width of the mass distribution) at position space. In contrast, Fisher's information decreases with the factor $a^{-2}$ at momentum space. An interesting result of the model emerged when we noticed that Fisher's information for our model is related to the uncertainties of position and momentum, namely, $F_{x}=4\sigma_{p}^{2} $ and $F_{p}=4\sigma_{x}^{2}$. It was also possible to conclude that for more localized mass distribution, the greater the transmitted information, i. e., the greater the position uncertainty, the smaller the model's momentum uncertainty.

Finally, we note that
\begin{equation}\label{rf}
\lim_{x\rightarrow\infty} \sech^{2}(ax)\cdot V(x)\to 0,
\end{equation}
i. e., the mass distribution seems to have a dominant characteristic in the system.  Therefore, the results presented here can be reproduced as long as the condition (\ref{rf}) is preserved.


\section*{Acknowledgments}
The author thank the Coordena\c{c}\~{a}o de Aperfei\c{c}oamento de Pessoal de N\'{i}vel Superior (CAPES), and Conselho Nacional de Desenvolvimento Cient\'{i}fico e Tecnol\'{o}gico (CNPq) for financial support. The author is grateful to C. A. S. Almeida, and M. S. Cunha for important discussions.

\end{document}